\pdfoutput=1
\documentclass[twocolumn]{aastex631}

\usepackage{graphicx}
\usepackage{threeparttable}
\usepackage{booktabs}
\usepackage{amsmath} 
\usepackage{amssymb}
\newcommand{\footnoterule}{
  \kern -1pt
  \hrule width 0.25\textwidth height 0.5pt
  \kern 1pt
}
\begin{document}
\title{A Series of (Net) Spin-down Glitches in PSR J1522$-$5735: Insights from the Vortex Creep and Vortex Bending Models}
\author[0000-0003-1480-2349]{S.Q. Zhou}
\affiliation{School of Physics and Astronomy, Sun Yat-Sen University, Zhuhai, 519082, China}
\author[0000-0002-1662-7735]{W.T. Ye}
\affiliation{Key Laboratory of Particle Astrophysics, Institute of High Energy Physics, Chinese Academy of Sciences, Beijing, 100049, China; \href{mailto:gemy@ihep.ac.cn}{gemy@ihep.ac.cn}}
\affiliation{University of Chinese Academy of Sciences, Chinese Academy of Sciences, Beijing 100049, China}
\author[0000-0002-3776-4536]{M.Y. Ge}
\affiliation{Key Laboratory of Particle Astrophysics, Institute of High Energy Physics, Chinese Academy of Sciences, Beijing, 100049, China; \href{mailto:gemy@ihep.ac.cn}{gemy@ihep.ac.cn}}
\affiliation{University of Chinese Academy of Sciences, Chinese Academy of Sciences, Beijing 100049, China}
\author{E. G\"{u}gercino\u{g}lu}
\affiliation{National Astronomical Observatories, Chinese Academy of Sciences, 20A Datun Road, Chaoyang District, Beijing, 100101, China; \href{mailto:egugercinoglu@gmail.com}{egugercinoglu@gmail.com}}
\author[0000-0003-2256-6286]{S.J. Zheng}
\affiliation{Key Laboratory of Particle Astrophysics, Institute of High Energy Physics, Chinese Academy of Sciences, Beijing, 100049, China; \href{mailto:gemy@ihep.ac.cn}{gemy@ihep.ac.cn}}
\affiliation{University of Chinese Academy of Sciences, Chinese Academy of Sciences, Beijing 100049, China}
\author[0000-0003-0454-7890]{C. Yu}
\affiliation{School of Physics and Astronomy, Sun Yat-Sen University, Zhuhai, 519082, China}
\author[0000-0002-5381-6498]{J.P. Yuan}
\affiliation{Xinjiang Astronomical Observatory, Chinese Academy of Sciences, Xinjiang, 830011, China}
\author{J. Zhang}
\affiliation{Department of Physics and Electronic Engineering, QiLu Normal University, Jinan, 250033, China}

\begin{abstract}
Through a detailed timing analysis of \textit{Fermi}-LAT data, the rotational behavior of the $\gamma$-ray pulsar PSR J1522$-$5735 was tracked from August 2008 (MJD 54692) to January 2024 (MJD 60320). During this 15.4-year period, two over-recovery glitches and four anti-glitches were identified, marking a rare occurrence in rotation-powered pulsars (RPPs). The magnitudes of these (net) spin-down glitches were determined to be $|\Delta\nu_{\rm g}/\nu| \sim 10^{-8}$, well above the estimated detectability limit. For the two over-recovery glitches, the respective recovery fractions $Q$ are $2.1(7)$ and $1.4(2)$. Further analysis showed no substantial variations in either the flux or pulse profile shape in any of these events, suggesting that small (net) spin-down glitches, unlike large events observed in magnetars and magnetar-like RPPs, may occur without leaving an impact on the magnetosphere. Within the framework of the vortex creep and vortex bending models, anti-glitches and over-recoveries indicate the recoupling of vortex lines that moved inward as a result of a crustquake; meanwhile, the apparent fluctuations in the spin-down rate after the glitches occur as a result of the coupling of the oscillations of bent vortex lines to the magnetosphere.
\end{abstract}

\keywords{\href{http://astrothesaurus.org/uat/628}{$\gamma$-ray astronomy (628)}; \href{http://astrothesaurus.org/uat/1408}{Rotation powered pulsars (1408)}; \href{http://astrothesaurus.org/uat/1916}{Time series analysis (1916)}}

\section{Introduction} \label{sec:intro}
Investigating pulsar glitches provides invaluable insights into the intricate internal dynamics and structure of neutron stars (for a review, see e.g., \citealt{antonelli22,Antonopoulou2022RPPh, Zhou2022Univ}), as well as associated phenomena such as gravitational wave emissions \citep{Haskell2023NatAs} and fast radio bursts (FRBs) \citep{Younes2023NatAs, Ge2024RAA, Hu2024Natur}. Among the plethora of telescopes worldwide, the \textit{Fermi} Large Area Telescope (\textit{Fermi}-LAT), operational since 2008, is a powerful instrument for studying glitches \citep{Ray2011ApJS, Abdo2013ApJS}, notably timing more than 290 $\gamma$-ray pulsars \citep{Smith2023ApJ}. The rotation-powered pulsar (RPP), PSR J1522$-$5735 (period $P=204\ \rm ms$ and period derivative $\dot{P}=62.46\times10^{-15}\ \rm s\ s^{-1}$), potentially associated with the supernova remnant SNR G321.9$-$0.3, is also a $\gamma$-ray source, which was discovered in blind searches of \textit{Fermi}-LAT data, with no radio pulsations detected in the Parkes follow-up observations \citep{Pletsch2013ApJ, Smith2023ApJ}. Importantly, an atypical glitch was observed in this radio-quiet $\gamma$-ray RPP \citep{Pletsch2013ApJ}.

Typical glitches display an interesting characteristic during the steady spin-down process of pulsars, wherein the spin frequency $\nu$ undergoes an instantaneous (on the order of seconds) increase ($\Delta\nu_{\rm g}>0$) and the spin-down rate $\dot\nu$ often decreases ($\Delta\dot\nu_{\rm g}\leq0$), followed by a recovery phase \citep{Espinoza2011MNRAS,Palfreyman2018Nature}. In the post-glitch recovery phase, $\nu$ and $\dot\nu$ either return to their initial pre-glitch rotation states or sustain permanent changes, sometimes initiating with exponential decay(s) \citep{Erbil2022MNRAS}. On the other hand, a small fraction of glitches exhibit atypical behaviors, classified as delayed \citep{Shaw2021MNRAS}, slow \citep{Zhou2019Ap&SS}, anti- \citep{Archibald2013Natur}, and over-recovery glitches \citep{Livingstone2010ApJ}, the latter two of which are specifically referred to as ``(net) spin-down'' glitches \citep{Archibald2017ApJ, Younes2023NatAs}. In PSR J1522$-$5735, the event that observed as a net spin-down at the completion of exponential recovery was identified as an over-recovery glitch \citep{Pletsch2013ApJ}.

The fact that pulsar emission remains unaltered during glitches indicates that these events originate intrinsically within the neutron star \citep{Fuentes2017A&A,zubieta24}, and several models have been proposed \citep{Zhou2022Univ}, including the widely accepted vortex creep model \citep{Alpar1984ApJ, Erbil2020MNRAS}. Capable of explaining typical glitch behaviors, the vortex creep model also provides estimates for the moments of inertia of the superfluid regions involved, the coupling time scales between various stellar components, and the epochs of future glitches \citep{Alpar1984ApJ, Erbil2019MNRAS, Erbil2022MNRAS}. More recently, \cite{Erbil2023MNRAS} have extended the vortex creep theory to the case of time-dependent magnetospheric changes \citep{erbil17} and incorporated the vortex bending oscillations to account for the low amplitude variations in $\dot\nu$. Despite significant advances in modeling typical glitches, the physical understanding of atypical glitches remains to be understood, and the vortex creep model has yet to fully explore their peculiar properties.

After \cite{Pletsch2013ApJ}, the \textit{Fermi}-LAT has accumulated more than a decade of regular timing observations on PSR J1522$-$5735. It is therefore highly worthwhile to examine its spin evolution during this extended time frame. In this work, we highlight the breakdown in the rotational evolution of this pulsar, attributed to multiple (net) spin-down glitches. Section \ref{sec:observations} provides an overview of the \textit{Fermi}-LAT observations of PSR J1522$-$5735 and the classic approach for identifying glitches. Glitch properties and emission variability analysis are detailed in Sections \ref{sec:results}. Section \ref{sec:model} explores the application of the vortex creep and the vortex bending models to these glitches. Finally, a summary is presented in Section \ref{sec:summary}.

\section{Observations and Data Analysis}\label{sec:observations}
Since August 2008, the \textit{Fermi} $\gamma$-ray Space Telescope collaboration, utilizing the Large Area Telescope (LAT), has been conducting a survey of the entire sky across the energy range from 20 MeV to 300 GeV (see \citealt{Abdo2013ApJS} for details). This work performed a detailed analysis of the \textit{Fermi}-LAT observations taken from August 2008 to January 2024 (MJDs 54692 to 60320) with the goal of tracking the rotational properties of PSR J1522$-$5735. For this purpose, the optimal method employed is ``pulsar timing", a diagnostic technique to evaluate the consistency between the observed times of arrival (ToAs) of pulsar pulses and the predictions from a truncated Taylor series phase model \citep{Edwards2006MNRAS}:
\begin{equation}
    \label{timingmodel}    
    \phi(t) = \phi_0 + \nu_0 (t - t_0) + \frac{1}{2!}\dot{\nu}_0 (t - t_0)^2 + \frac{1}{3!}\ddot{\nu}_0 (t - t_0)^3,
\end{equation}
where $\phi_0$, $\nu_0$, $\dot{\nu}_0$, and $\ddot{\nu}_0$ are the phase, spin-frequency and its derivatives, measured at the reference epoch $t_{0}$. The differences between the observed and the predicted ToAs are known as ``timing residuals". In the context of a glitch, timing residuals deviate from zero because the model (\ref{timingmodel}) does not account for an additional phase, which involves a permanent step change in $\phi$, $\nu$, and $\dot\nu$ (denoted by a subscript p) at the glitch epoch, along with potential multiple exponentially relaxing components with decay timescales (denoted by a subscript d) \citep{Yuan2010ApJ, Ge2020ApJa}:
\begin{equation}
    \label{glitchmodel}
\begin{split}
    \phi_{\rm g} (t) = &\Delta\phi + \Delta \nu_{\rm p} (t - t_{\rm g}) + \frac{1}{2!}\Delta \dot{\nu}_{\rm p} (t - t_{\rm g})^2 +\\
    & \sum_{i}^N \Delta \nu_\mathrm{d}^{(i)} \tau_\mathrm{d}^{(i)} [ 1 - e^{-(t - t_{\rm g})/\tau_\mathrm{d}^{(i)}} ].
\end{split}
\end{equation}

From this, the occurrence of glitches and the evolution of both \(\nu\) and \(\dot{\nu}\) are determinable \citep{Ge2020ApJb}. At the time of a glitch, the total \(\nu\) and \(\dot{\nu}\) changes are \(\Delta\nu_{\rm g} = \Delta\nu_{\rm p} + \sum_{i}^N \Delta \nu_\mathrm{d}^{(i)}\) and \(\Delta\dot{\nu}_{\rm g} = \Delta\dot{\nu}_{\rm p} - \sum_{i}^N (\Delta \nu_\mathrm{d}^{(i)}/\tau_\mathrm{d}^{(i)})\), respectively \citep{Yuan2010MNRAS}. A parameter \(Q \equiv \sum_{i}^N \Delta \nu_\mathrm{d}^{(i)}/\Delta\nu_{\rm g}\) is often used to quantify the recovery degree of glitches. To obtain ToAs at the solar system barycenter from \textit{Fermi}-LAT observations using standard Fermi Science Tools (\texttt{v10r0p5}), the procedure described in \cite{Ge2019NatAs} was followed. The resulting ToAs from every 10 days exposure were found to be adequate for performing timing analysis on this pulsar.

\begin{figure*}
    \centering
    \includegraphics[width=0.88\textwidth]{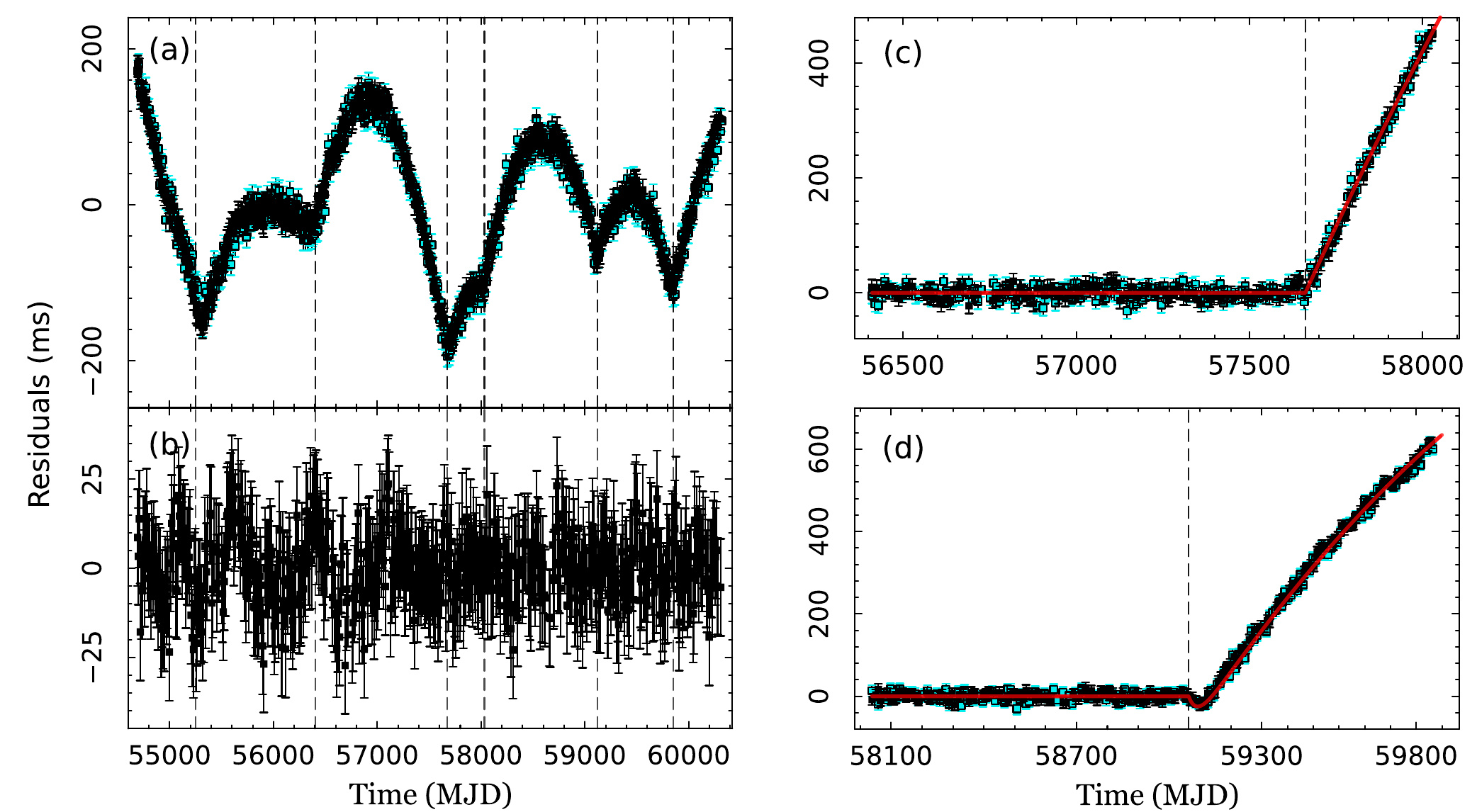}
    \caption{Panels (a) and (b) present timing residuals after fitting for the pulsar parameters and a second frequency derivative $\ddot\nu$ over the span of the data; however, Panel (b) additionally includes the glitch model (Eq. (\ref{glitchmodel})) in the fitting. Panels (c) and (d) show timing residuals relative to a pre-glitch, spin-down model (including  $\nu$, $\dot\nu$) for anti-glitch G3 (similar to G2, G4, G6)  and over-recovery glitch G5  (similar to G1), respectively. In Panels (a), (c), and (d), residuals in black correspond to \textit{Fermi}-LAT observations, while those in cyan indicate simulated observations generated using the \texttt{TEMPO2} \texttt{fake} plugin with parameters in Table \ref{tab:glitch}. Red lines in Panels (c) and (d)  represent the polynomial detrending of the residuals based on an analytical glitch model (Eq. (\ref{glitchmodel})), and vertical lines across all panels mark the epochs of glitches.}
    \label{fig:residuals}
\end{figure*}
\begin{deluxetable*}{lcccccc}
    \tabletypesize{\footnotesize}
    \tablewidth{1\textwidth} 
    \tablecaption{Pulsar and glitch parameters of PSR J1522$-$5735, including glitch magnitude revisions (with superscript "rev``) after accounting for the $\dot\nu$ oscillations in the framework of the vortex bending model (see Table \ref{tab:vortex_bending_model} below).}
    \tablecolumns{7}
    \tablehead{Parameter & \multicolumn{6}{c}{PSR J1522$-$5735}}
    \startdata
    Right ascension, $\alpha$ (J2000.0)\dotfill & \multicolumn{6}{c}{$15^{\rm h}22^{\rm m}05^{\rm s}.30(5)$} \\
    Declination, $\delta$ (J2000.0)\dotfill & \multicolumn{6}{c}{$-57{\degr}35{\arcmin}00{\farcs}0(3)$} \\
    Epoch, $t_{0}$ (MJD) \dotfill & \multicolumn{6}{c}{54969} \\
    $\nu_0$ (Hz) \dotfill & \multicolumn{6}{c}{4.8954708041(4)} \\
    $\dot\nu_0$ ($10^{-12}\ \rm Hz\ s^{-1}$)\dotfill & \multicolumn{6}{c}{$-$1.49685(7)} \\
    $\ddot\nu_0$ ($10^{-24}\ \rm Hz\ s^{-2}$)\dotfill & \multicolumn{6}{c}{$-$5.0(8)} \\
    Characteristic age, $\tau_{c}$ (kyr)\dotfill & \multicolumn{6}{c}{51.8} \\
    Spin-down power, $\dot{E}$ ($10^{34}\ \rm erg\ s^{-1}$)\dotfill & \multicolumn{6}{c}{28.9} \\
    Surface magnetic field, $B_{S}$ ($10^{12}\ \rm G$)\dotfill & \multicolumn{6}{c}{3.61} \\
    \hline
    Glitch No.\dotfill & G1\hyperref[note]{*} & G2 & G3 & G4 & G5 & G6 \\
    \hline
    Glitch epoch, $t_{\mathrm g}$ (MJD)\dotfill & 55251(5) & 56405(6) & 57661(4) & 58032(6) & 59063(4) & 59852(4) \\
    $\Delta\nu_{\rm p}$ ($10^{-9}\ \rm Hz$)\dotfill& $-$55(2) & $-$50.5(8) & $-$70(4) & $-$41(3) & $-$61(1) & $-$68(3) \\
    $\Delta\dot\nu_{\rm p}$ ($10^{-15}\ \rm Hz\ s^{-1}$)\dotfill & 1.15(9) & 0.93(9) & 0.5(2) & 0.2(2) & 0.81(7) & $-$0.5(1) \\
    $\Delta\nu_{\rm d}$ ($10^{-9}\ \rm Hz$)\dotfill& 107(31) & -- & -- & -- & 208(74) & -- \\
    $\tau_{\rm d}$ (d)\dotfill & 45(14) & -- & -- & -- & 22(6) & -- \\
    $Q$\dotfill & 2.1(7) & -- & -- & -- & 1.4(2) & -- \\
    $\Delta\nu_{\rm g}/\nu$ ($10^{-9}$) \dotfill& 11(6) & $-$10.3(2) & $-$14.2(7) & $-$8.4(7) & 30(15) & $-$13.9(6) \\
    $\Delta\dot\nu_{\rm g}/\dot\nu$ ($10^{-3}$)\dotfill & 17(8) & $-$0.62(6) & $-$0.4(1) & $-$0.1(1) & 72(33) & 0.30(9) \\
    Data span (MJD) \dotfill & 54692$-$56400 & 55255$-$57658 & 56410$-$58027 & 57665$-$59060 & 58038$-$59849 & 59067$-$60320 \\
    Number of ToAs \dotfill & \multicolumn{6}{c}{542} \\
    Pre-/post-fit RMS ($\rm \mu s$) \dotfill & \multicolumn{6}{c}{77663/9377} \\
    \hline
    $\Delta\nu_{\rm p}^{\rm (rev)}$ ($10^{-9}\ \rm Hz$)\dotfill& $-62(7)$ & $-$43(5) & $-$72(22) & $-$43(36) & $-55(21)$ & $-$64(13) \\
    $\Delta\dot\nu_{\rm p}^{\rm (rev)}$ ($10^{-15}\ \rm Hz\ s^{-1}$)\dotfill & 0.98(25) & 0.65(19) & 0.94(380) & 0.14(590) & 0.29(51) & $-$0.98(65) \\
    $\Delta\nu_{\rm d}^{\rm (rev)}$ ($10^{-9}\ \rm Hz$)\dotfill& 11.5(65) & -- & -- & -- & 25(8) & -- \\
    $\tau_{\rm d}^{\rm (rev)}$ (d)\dotfill & 9(4) & -- & -- & -- & 11(2) & -- \\
    $Q^{\rm (rev)}$\dotfill & $-0.23(16)$ & -- & -- & -- & $-0.85(78)$ & -- \\
    $\Delta\nu_{\rm g}/\nu ^{\rm (rev)}$ ($10^{-9}$) \dotfill& $-10.3(20)$ & $-$8.7(11) & $-$14.6(44) & $-$8.7(74) & $-6.1(46)$ & $-13(3)$ \\
    $\Delta\dot\nu_{\rm g}/\dot\nu^{\rm (rev)}$ ($10^{-3}$)\dotfill & 9.5(28) & $-$0.4(1) & $-$0.63(250) & $-$0.1(39) & 17(4) & 0.6(4) \\
\enddata
\begin{tablenotes}
    \item[] \textbf{Notes}. The position is adopted from \cite{Smith2023ApJ}. Uncertainties, indicated within parentheses as 1-$\sigma$, were derived using \texttt{TEMPO2}. $^*$\label{note}Following \cite{Smith2023ApJ}, corrections have been made to the erroneous pulsar parameters and G1 results initially reported by \cite{Pletsch2013ApJ}.\\
\end{tablenotes}
    \label{tab:glitch}
\end{deluxetable*}
\section{Over-recovery glitches and Anti-glitches of PSR J1522$-$5735} \label{sec:results}
\begin{figure*}
    \centering
    \includegraphics[width=0.88\textwidth]{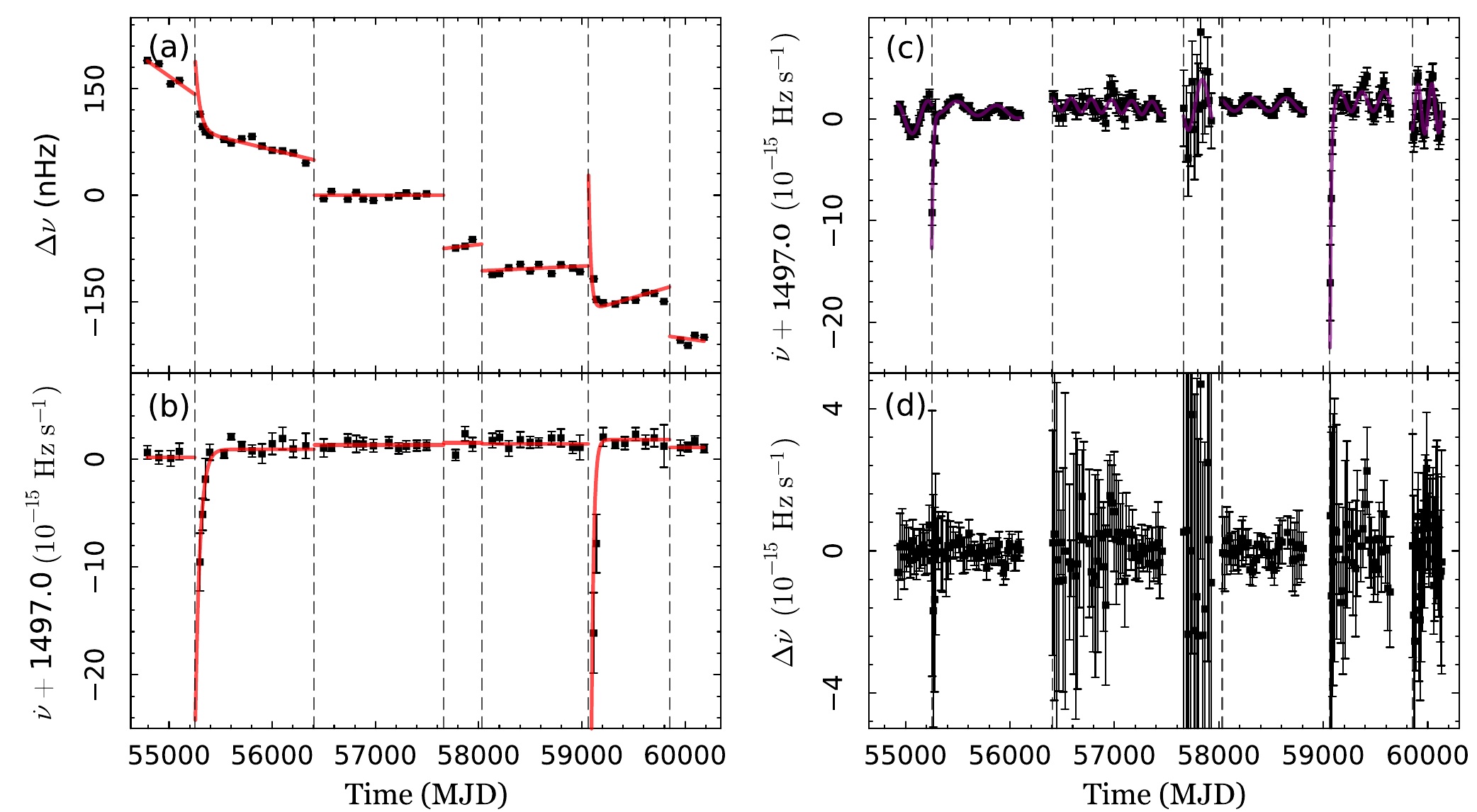}
    \caption{Panel (a) presents frequency residuals $\Delta\nu$ relative to the inter-glitch G2-G3 evolution trend. Panel (b) shows the time-dependent evolution of $\dot\nu$. Panel (c) displays the fit of the vortex bending model (Eq. (\ref{eq:bending})) to the inter-glitch oscillatory $\dot\nu$, while Panel (d) highlights discrepancies between the fitted model and and the observations. Measurements of $\nu$ and $\dot\nu$ were obtained through a partially phase-coherent analysis after dividing the 15.4-year dataset into 200-day subsets. In these measurements, the overlap of data used in Panel (c) is significantly greater than that in Panel (b), thereby enhancing the visibility of $\dot\nu$ oscillations. Panels (a) and (b) present the values at the mid-epoch, whereas Panel (c) displays the values at the left-epoch for each subset. Red lines in Panels (a) and (b) represent rotational evolution based on an analytical glitch model (Eq. (\ref{glitchmodel})), and purple lines in Panel (c) are derived from the vortex bending model (Eq. (\ref{eq:bending})); vertical lines across all panels mark the epochs of glitches.}
    \label{fig:evolution}
\end{figure*}

Implementing the aforementioned method with the popular pulsar timing package \texttt{TEMPO2} \citep{Hobbs2006MNRAS} yields timing residuals for PSR J1522$-$5735, enabling the visual identification of six (net) spin-down glitches: two over-recovery glitches (G1, G5) and four anti-glitches (G2, G3, G4, G6). In order to determine the parameters defined in Eqs.(\ref{timingmodel}) and (\ref{glitchmodel}), each glitch epoch was consistently estimated by averaging ToAs $T1$ and $T2$ that surround the respective glitch, with the error calculated as $(T2-T1)/2$. The timing model, inclusive of glitches, was directly applied to the ToAs across the entire dataset using least-squares fitting to minimize the root-mean-square (RMS) value of the residuals. The fitting results are summarized in Table \ref{tab:glitch}, with Panels (a) and (b) of Figure \ref{fig:residuals} showing the timing residuals both pre- and post-fitting of glitches. The evolution of $\nu$ and $\dot\nu$ spanning 15.4 years is shown in Panels (a) and (b) of Figure \ref{fig:evolution}.  

Over-recovery glitches, G1 and G5, are identified through a characteristic negative dip followed by a rise toward positive values in timing residuals, as illustrated in Panel (d) of Figure \ref{fig:residuals}  — similar to the pattern of the over-recovery glitch in magnetar AXP 4U 0142+61 \citep{Gavriil2011ApJ}. G1, as reported by \cite{Pletsch2013ApJ}\hyperref[note]{*}, and G5 show initial jumps in spin parameters ($\Delta\nu_{\mathrm g}>0$ and $\Delta\dot\nu_{\mathrm g}<0$) but quickly decay, overshooting their values ($\Delta\nu_{\mathrm p}<0$ and $\Delta\dot\nu_{\mathrm p}>0$), with respective recovery fractions $Q$ of $2.1(7)$ and $1.4(2)$. To date, glitches featuring high $Q$ recoveries have been observed in both young and old pulsars, yet glitches with $Q>1$ have been reported only for three sources, in total five instances: twice in magnetar AXP 4U 0142+61 \citep{Gavriil2011ApJ, Archibald2017ApJ}, twice in magnetar-like high magnetic field RPP J1119$-$6127 \citep{Antonopoulou2015MNRAS}, and once in another magnetar-like high magnetic field RPP J1846$-$0258 \citep{Livingstone2010ApJ}. Unlike these young pulsars, over-recovery glitches in middle-aged PSR J1522$-$5735 have values for $\Delta\nu_{\rm g}/\nu$ and $\Delta\dot\nu_{\rm g}/\dot\nu$ that are several orders of magnitude smaller.

Anti-glitches, G2, G3, G4 and G6, unmistakably correspond to a linear drift toward positive values in timing residuals (see the Panel (c) of Figure \ref{fig:residuals}), significantly distinguishing them from behaviors observed in over-recovery glitches, G1 and G5. Further analysis has revealed that these glitches are of similar magnitude, with $\Delta\nu_{\rm g}/\nu\sim-1\times10^{-8}$ and no appreciable relaxation. G2, G3, and G4 exhibit the typical anti-glitch characteristic (see Table \ref{tab:glitch} and the Panels (a) and (b) of Figure \ref{fig:evolution}), defined as having $\Delta\nu_{\rm g} < 0$ and $\Delta\dot\nu_{\rm g}\geq 0$ \citep{Espinoza2021A&A}. It is important to note that in scenarios where the relaxation time is shorter than the observation interval around anti-glitch epoch, it becomes challenging to exclude the possibility that these events are the net effect following the recovery of decaying component. Currently, observations of anti-glitches have been limited to RPP PSR B0540$-$69 \citep{tuo24}, accreting pulsar NGC 300 ULX-1 \citep{Ray2019ApJ}, as well as magnetars SGR 1935+2154 \citep{Younes2023NatAs} and 1E 2259+586 \citep{Younes2020ApJ}. The discovery of anti-glitches for the first time in a radio-quiet $\gamma$-ray RPP makes this phenomenon even more intriguing.

With $|\Delta\nu_{\rm g}|<1\rm\ \mu Hz$ and $|\Delta\nu_{\rm g}/\nu|<50\times10^{-9}$, the detected events fall into the category of small glitches \citep{Espinoza2011MNRAS, Espinoza2014MNRAS}. The identification of small glitches can be perturbed by timing noise, a widespread phenomenon in pulsar rotation that manifests as small-amplitude changes in $\nu$ relative to the spin-down model, as noted in current studies (e.g., \citealt{Chukwude2010MNRAS, Espinoza2011MNRAS, Espinoza2014MNRAS, Espinoza2021A&A, grover24}). Nevertheless, multiple (net) spin-down glitch signatures in PSR J1522$-$5735's timing residuals are clearly distinguishable from timing noise. \cite{Espinoza2014MNRAS} suggested a simple way to assess glitch detectability limits for a given pulsar. For PSR J1522$-$5735, substituting $\Delta T=10\ \mathrm{days}$ and $\sigma_{\phi} = 0.05\ \mathrm{rotations}$ into equation (2) in \cite{Espinoza2014MNRAS} shows that all detected glitches significantly exceed the minimum detectable glitch size.  As a validation of the results for PSR J1522$-$5735, the \texttt{fake} plugin for \texttt{TEMPO2} was employed with glitch parameters to simulate ToAs, while ensuring that the observational cadence and ToA errors aligned with those in \textit{Fermi}-LAT observations. The comparison shown in the Panels (a), (c), and (d) of Figure \ref{fig:residuals} between the simulated and the observed residuals implies that our measurements of glitches can effectively describe anomalies in the timing data. Furthermore, the inclusion of glitch parameters in the model greatly improves the fitting of residuals (see the Panel (b) of Figure \ref{fig:evolution}), as evidenced by the significantly reduced RMS value. Ultimately, we are optimistic that PSR J1522$-$5735 is the first observed normal RPP experiencing both anti- and over-recovery glitches. 

Given that every recorded (net) spin-down glitch, except for two small cases \citep{Antonopoulou2015MNRAS, tuo24}, has been linked to radiative changes such as flux enhancements or pulse shape changes \citep{Hu2019AN, Hu2023ApJ}, this analysis extends to examining whether PSR J1522$-$5735 exhibits a similar connection. No significant changes were detected in $\gamma$-ray flux or pulse profile morphology before and after the glitches\footnote{\url{https://fermi.gsfc.nasa.gov/ssc/data/access/lat/10yr_catalog/ap_lcs.php?ra=14-15}}. This suggests that (net) spin-down glitches may occur independently of the magnetosphere, particularly in smaller events. Another possibility is that glitch induced ephemeral changes are taking place close to the polar cap of PSR J1522$-$5735 with small distortions in the magnetic field lines affecting radio emission \citep{akbal15,Zhou2023MNRAS}, but hindered in a radio-quiet RPP \citep{Erbil2022MNRAS}.
\section{Insights from the vortex creep and the vortex bending models} \label{sec:model}

\begin{figure}
    \centering
    \includegraphics[width=1.0\linewidth]{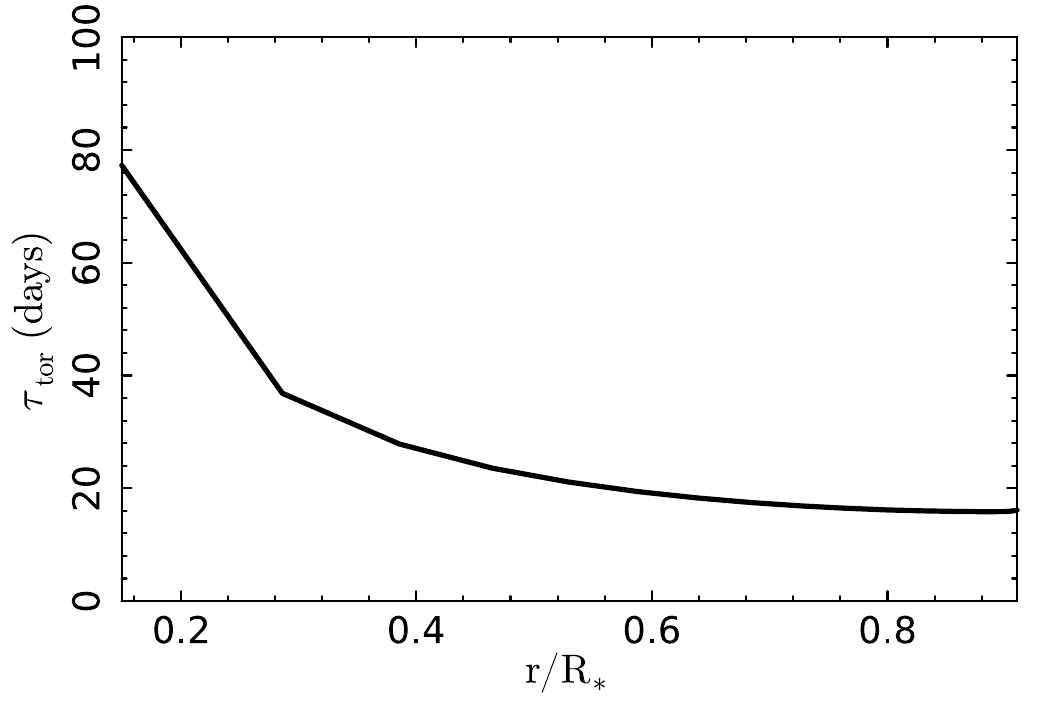}
    \caption{Exponential decay timescale predicted for PSR J1522--5735 by the vortex creep model [Model 3 in \cite{Erbil2017MNRAS}] versus the distance from the stellar center normalized to radius.}
    \label{tautor}
\end{figure}

Next, the vortex creep and bending models are applied to explore  the properties of (net) spin-down glitches of PSR J1522$-$5735. For the exponential recoveries following G1 and G5, we investigated the possibility of the neutron star's core response to the sudden spin-ups. According to the vortex creep model, vortex lines' creep across the toroidally oriented flux tubes contributes an exponentially decaying component in the post-glitch relaxation \citep{Erbil2017MNRAS}. The associated relaxation time is given by \citep{erbil14}
\begin{equation}
\resizebox{0.97\hsize}{!}{$
\begin{aligned}
    \tau_{\rm tor} \simeq & \ 60 \left(\frac{\vert\dot{\Omega}\vert}{10^{-10}\,\mbox{rad s$^{-1}$}}\right)^{-1} \left(\frac{T}{10^{8}\,\mbox{K}}\right) \left(\frac{R}{10^{6}\,\mbox{cm}}\right)^{-1} x_{\rm p}^{1/2} \\
    & \times \left(\frac{m_{\rm p}^*}{m_{\rm p}}\right)^{-1/2} \left(\frac{\rho}{10^{14}\,\mbox{g cm$^{-3}$}}\right)^{-1/2} \left(\frac{B_{\phi}}{10^{14}\,\mbox{G}}\right)^{1/2} \, \mbox{days,}  
\end{aligned}
$}
\label{eq:tautor}
\end{equation}
where $\dot{\Omega}$ is the spin-down rate, $T$ is the neutron star interior temperature, $x_{\rm p}$ is the proton fraction in the neutron star core, $\rho$ is the matter density, $ m_{\rm p}^*(m_{\rm p})$ is the effective (bare) mass of  protons, and $B_{\phi}$ is the strength of the toroidal component of the magnetic field. The relaxation time-scale $\tau_{\rm tor}$ associated with the toroidal field region in the core strongly depends on the equation of state (EOS) of the core matter in a neutron star. As a representative, here we only consider a 1.4 solar mass (1.4M$_{\odot}$) neutron star model, Model 3 in \cite{Erbil2017MNRAS}. We take the matter density $\rho$ and the other EOS related physical parameters, i.e. the proton fraction $x_{\rm p}$ and the distance from the center normalized to the neutron star radius $r/R_{*}$ corresponding to SLy4 EOS from \cite{Douchin2001}. We use the effective to bare mass ratio for protons $m^{*}_{\rm p}/m_{\rm p}$ as a function of the matter density in the core as given in \cite{Chamel2008}. As for the interior temperature $T$, we use the value predicted by the standard cooling of a neutron star via modified Urca process \citep{Yakovlev2011}, while for the toroidal field strength we employ the prescription given by $B_{\phi}\approx10^{14}(B_{\rm p}/10^{12} \mbox{G})^{1/2}$ G (with $B_{\rm p}$ being the dipolar field strength at the magnetic pole) in \cite{erbil14}.
The decay time-scale $\tau_{\rm tor}$ associated with the toroidal flux region for Model 3 in \cite{Erbil2017MNRAS} is applied to PSR J1522$-$5735, and the result is shown in Figure \ref{tautor}. As can be clearly seen from the same figure, $\tau_{\rm tor}$ agrees qualitatively well with the observed decay time-scales in Table \ref{tab:glitch}. 

Anti-glitches and over recoveries, on the other hand, may be easily achieved by inward transportation of the vortex lines via broken platelet in a quake and recoupling of the corresponding lines with the external braking torque \citep{akbal15,erbil17}. When the vortex lines are  transported radially inward by an externally driving force like quakes, the superfluid recieves angular momentum from the crust, thereby slows down the observed surface of a neutron star during a glitch event. The response of the spin of a neutron star to a perturbation of the distribution of its vortices by such an inward motion would be a delayed and enhanced outward vortex flow by which over-relaxation in the spin-down rate ensues \citep{akbal15,erbil17}. The recent numerical simulations supports this view \citep{Howitt2022MNRAS}.

\begin{deluxetable*}{lccccccc} 
    \label{tab:vortex_bending_model}
    \tablewidth{1.0\textwidth} 
    \tabletypesize{\footnotesize}
    \tablecaption{Parameters of the vortex bending model for inter-glitch oscillatory $\dot\nu$.}
    \tablecolumns{8}
    \tablehead{Inter-glitch & Pre-G1 & G1-G2 & G2-G3 & G3-G4 & G4-G5 & G5-G6 & Post-G6}
    \startdata
    $A$ ($10^{-8} \rm\ rad\ s^{-1}$)\dotfill & 4.01(42) & 1.46(30) & 1.09(21) & 5.13(351) & 1.86(26) & 1.75(38) & 2.79(42) \\
    $\tau$ ($\rm d$)\dotfill & $-$ & 248 & $-$ & 135 & $-$ & $-$ & $-$ \\
    $\Omega_0$ ($10^{-7}\rm\ rad\ s^{-1}$)\dotfill & 2.41(35) & 1.81(29) & 3.76(102) & 2.91(233) & 2.20(49) & 3.45(110) & 5.47(105) \\
    $\phi$ (rad)\dotfill & $-$1.37(13) & $-$1.60(11) & 0.90(19) & $-$1.11(105) & $-$4.66(13) & $-$2.47(22) & 4.46(14) \\
    $\dot\nu_{0}$ ($10^{-15}\ \rm Hz\ s^{-1}$)\dotfill & $-$1496.82(15) & $-$1496.19(5) & $-$1495.78(9) & $-$1495.21(178) & $-$1495.57(7) & $-$1495.26(21) & $-$1495.99(26) \\
    $\ddot\nu_{0}$ ($10^{-24}\ \rm Hz\ s^{-2}$)\dotfill & 11(16) & $-$10(3) & $-$5(3) & $-$10(280) & $-$0.87(334) & 2(11) & 14(35) \\
    $\dot\nu_{\rm d}$ ($10^{-15}\ \rm Hz\ s^{-1}$)\dotfill & -- & $-$15(4) & -- & -- & -- & $-$25(6) & -- \\
    $\tau_{\rm d}$ ($\rm d$)\dotfill & -- & 9(4) & -- & -- & -- & 11(2) & -- \\
    Reduced $\chi^2$ \dotfill & 0.59 & 0.77 & 0.70 & 0.60 & 0.57 & 1.25 & 0.80 \\
    Braking index $n=\nu\ddot\nu/\dot\nu^{2}$\dotfill & 23(36) & $-$21(6) & $-$10(7) & $-$22(610) & $-$2(7) & 5(24) & 31(76) \\
    \enddata
\end{deluxetable*}

According to the vortex bending model \citep{Erbil2023MNRAS}, in order for vortex lines to pin to nuclei in the crust and flux tubes in the outer core they have to be bent over slightly but the finite tension of the lines resists and tends to bring the vortices back to the straight configuration. Such shape transitions induced by externally driven events like glitches, crustquakes, outbursts, pulsar wind ejection and giant pulses apply a time-variable superfluid torque acting on the surface of a neutron star as vortex lines  display low amplitude and low frequency oscillations about equilibrium configuration. The net result is damped sinusoidal oscillations seen in the spin-down rate of pulsars, and given by \citep{Erbil2023MNRAS} 
\begin{equation}
\begin{aligned}
    \dot{\nu}_{\rm c}(t) &= \frac{A}{2\pi} \Omega_0 \exp\left(-\frac{t}{2\tau}\right) \cos\left(\Omega_0 t + \phi\right) \\
    &\quad - \frac{A}{4\pi\tau} \exp\left(-\frac{t}{2\tau}\right) \sin\left(\Omega_0 t + \phi\right) + \dot{\nu}_0 + \ddot{\nu}_0 t,   
\end{aligned}
    \label{eq:bending}
\end{equation}
where $A$, $\phi$ and $\tau$ are the amplitude, phase and damping time-scale of the oscillations, respectively. Vortex bending over a length-scale of $\sim10^{5}$\,cm results in oscillations with a period $T=2\pi/\Omega_0$ of a few hundred days. When the spin-down rate of a pulsar becomes $|\dot\nu|\lesssim10^{-12}$\,Hz s$^{-1}$, the external pulsar braking torque and internal torque contribution due to vortex bending turn out to be of comparable magnitudes and oscillations in the spin-down rates for those pulsars are anticipated. The fit of the vortex bending model, Eq. (\ref{eq:bending}), to oscillatory $\dot\nu$ is presented in Panel (c) of Figure \ref{fig:evolution}, with the corresponding fitted values provided in Table \ref{tab:vortex_bending_model}; the residuals in $\dot\nu$, after accounting for the oscillations of vortex lines, are highlighted in Panel (d). An inspection of Table \ref{tab:vortex_bending_model} reveals that in some inter-glitch data segments the braking index $n$ is negative, contrary to expectations from slow-down via magnetic dipole radiation or wind \citep{zhang22}. Besides, after large glitches anomalously large, positive braking indices are anticipated due to decoupling of pinned crustal superfluid from the external torque during a spin-up event \citep{alpar06}. As also noted for PSR B0950+08 case, an increase in the dipolar magnetic field would result in negative $\ddot\nu$ and in turn negative $n$ \citep{Erbil2023MNRAS}. There may be several circumstances which potentially lead to strengthening of the dipolar component of magnetic field. PSR J1522$-$5735 is a very active glitching source, with six events in 15.4 years, and quite likely it had undergone many more glitch events in the past. An outcome of a series of glitches would be heating of the pulsar's crust via dissipation of either rotational energy by superfluid friction \citep{alpar06} or mechanical energy by crustquakes \citep{larson02}. Inside a warmer crust the magnetic field evolution would accelerate. There seems to be two viable mechanisms to realize. One possibility is that if the initial magnetic field of a pulsar is buried by fallback matter inherited from supernova explosion, during the course of evolution the magnetic field grows gradually by reduction in the conductivity \citep{ho15}. Another possibility is that Hall drift assisted magnetic field evolution effectively converts the magnetic energy of the toroidal component into the poloidal one, thereby supporting a temporarily increased dipolar magnetic field \citep{gourgouliatos15, Gao2017ApJ, Wang2020Univ}. Future observations will help to test and discriminate among these hypotheses.        
\section{Summary} \label{sec:summary}
The long-term spin evolution of the middle-aged ($\tau_{c}\sim51.8\ \rm kyr$) pulsar PSR J1522$-$5735 between August 2008 and January 2024 is nicely revealed through the \textit{Fermi}-LAT observations. Over this period, this pulsar frequently glitched; specifically, it experienced two over-recovery glitches and four anti-glitches. The sizes of six glitches were measured with $|\Delta\nu_{\rm g}/\nu|$ ranging from $8.4\times 10^{-9}$ to $30\times 10^{-9}$, smaller than any previously recorded event in magnetars and high magnetic field RPPs. Still, the observed timing behavior is highly anomalous for a normal RPP. Examination of the fluxes and $\gamma$-ray pulse profiles showed no changes coincident with the (net) spin-down glitches. The vortex creep model invoked to explain normal glitches does hold for the non-trivial scenario of PSR J1522$-$5735. Anti-glitches and over recoveries are naturally produced by vortex inward motion and affected coupling between the crustal superfluid and the pulsar braking torque after a glitch. Also, exponential recovery timescales can be reproduced by the response of vortex lines' creep against to flux tubes during a glitch (see Table \ref{tab:glitch} and Figure \ref{tautor}). Consideration of the vortex bending model's prediction regarding removing sinusoidal-like oscillations in the spin-down rate and its application to the data reduces the residuals and improves the timing solution, see the Panels (c) and (d) of Figure \ref{fig:evolution} and Table \ref{tab:glitch}. 

\begin{acknowledgments}
\smallskip
\noindent {\it \textbf{Acknowledgements:}}
The authors thank the \textit{Fermi} $\gamma$-ray Space Telescope (\textit{Fermi}) team for their efforts in collecting and releasing data to the public. This work is supported by the National Key R\&D Program of China (2021YFA0718500) from the Minister of Science and Technology of China (MOST). S.Q.Z. and C.Y. are funded by the National SKA Program of China (2022SKA0120101), the Sun Yat-Sen University Graduate Education Innovation Project 2024 (2024$\_$71000$\_$B24783), and the Sichuan Science and Technology Program (2024NSFSC0456). M.Y.G. is funded by the National Natural Science Foundation of China (12373051) and the International Partnership Program of the Chinese Academy of Sciences (113111KYSB20190020). E.G. is funded by the National Natural Science Foundation of China (NSFC) programme 11988101 under the foreign talents grant QN2023061004L. J.Z. is funded by the National Natural Science Foundation of China (12373109 and U2031121), the National Key Research and Development Program of China (SQ2023YFB4500096), and the Natural Science Foundation of Shandong Province (ZR2020MA063). 
\end{acknowledgments}

\bibliography{sample631}{}
\bibliographystyle{aasjournal}
\end{document}